\documentclass[prc,showpacs,preprintnumbers,
               superscriptaddress,amsmath,amssymb,floatfix]{revtex4}

\usepackage{dcolumn}
\usepackage{bm}
\usepackage{longtable}

\usepackage{graphicx,epsfig,latexsym,amssymb}
\usepackage{multirow,amsmath,array,booktabs,color}
\usepackage[section]{placeins}

\newcommand{\tm}{\times}

\begin{document}
\title{Real stabilization method for nuclear single particle resonances}

\author{Li Zhang}
 \affiliation{School of Physics, Peking University, Beijing 100871, China}

\author{Shan-Gui Zhou}
 \email[E-mail: ]{sgzhou@itp.ac.cn}
 \affiliation{Institute of Theoretical Physics, Chinese Academy of Sciences,
              Beijing 100080, China}
 \affiliation{Center of Theoretical Nuclear Physics, National
              Laboratory of Heavy Ion Accelerator, Lanzhou 730000, China}

\author{Jie Meng}
 \affiliation{School of Physics, Peking University, Beijing 100871, China}
 \affiliation{Institute of Theoretical Physics, Chinese Academy of Sciences,
              Beijing 100080, China}
 \affiliation{Center of Theoretical Nuclear Physics, National
              Laboratory of Heavy Ion Accelerator, Lanzhou 730000, China}

\author{En-Guang Zhao}
 \affiliation{Institute of Theoretical Physics, Chinese Academy of Sciences,
              Beijing 100080, China}
 \affiliation{School of Physics, Peking University, Beijing 100871, China}
 \affiliation{Center of Theoretical Nuclear Physics, National
              Laboratory of Heavy Ion Accelerator, Lanzhou 730000, China}

\begin{abstract}
We develop the real stabilization method within the framework of the
relativistic mean field (RMF) model. With the self-consistent
nuclear potentials from the RMF model, the real stabilization method
is used to study single-particle resonant states in spherical
nuclei. As examples, the energies, widths and wave functions of
low-lying neutron resonant states in $^{120}$Sn are obtained. These
results are compared with those from the scattering phase shift
method and the analytic continuation in the coupling constant
approach and satisfactory agreements are found.
\end{abstract}

\pacs{02.60.Lj, 21.10.-k, 21.60.-n, 25.70.Ef}

\maketitle

\section{Introduction}
\label{sec:intro}

The investigation of continuum and resonant states is an important
subject in quantum physics. In recent years, there has been an
increasing interest in the exploration of nuclear single particle
states in the continuum. The construction of the radioactive ion
beam facilities makes it possible to study exotic nuclei with
unusual N/Z ratios. In these nuclei, the Fermi surface is usually
close to the particle continuum, thus the contribution of the
continuum and/or resonances being essential for exotic nuclear
phenomena~\cite{Bulgac80, Dobaczewski84, Dobaczewski96, Meng96,
Poeschl97, Meng98}. It has been also revealed that the contribution
of the continuum to the giant resonances mainly comes from
single-particle resonant states~\cite{Curutchet89, Cao02}.

For the theoretical determination of resonant parameters (the energy
and the width), 
several bound-state-like methods have been developed. The complex
scaling method (CSM) describes the discrete bound and resonant
states on the same footing~\cite{Kato06}. In this method, a complex
coordinate scaling is introduced to rotate the continuum into the
complex energy plane and the wave functions of resonant states, but
not scattering states, are transformed into square-integrable
functions~\cite{Kruppa88}. Although it involves the solution of a
complex eigenvalue problem which causes some difficulties in
practice, the CSM has been widely and successfully used to study
resonances in atomic and molecular systems~\cite{Reinhardt82, Ho83,
Moiseyev98} and atomic nuclei~\cite{Gyarmati86, Kruppa88, Kruppa97,
Arai06, Kato06}. The analytical continuation in the coupling
constant (ACCC) approach is based on an intuitive idea that a
resonant state can be lowered to be bound when the potential becomes
more attractive or equivalently the coupling constant stronger, thus
a resonant state being related to a series of bound states via an
analytical continuation in the coupling constant~\cite{Kukulin77,
Kukulin79, Kukulin89}. Combined with the cluster model, the ACCC
approach has been used to calculate the resonant energies and widths
in some light nuclei~\cite{Tanaka97, Tanaka99}. An attempt to
explore the unbound states by the ACCC approach within relativistic
mean field (RMF) model was first made in Ref.~\cite{Yang01} where
resonant parameters of some low lying resonant states in $^{16}$O
and $^{48}$Ca obtained from the ACCC calculations are comparable
with available data. The wave functions of nuclear resonant states
were also determined by the ACCC method where the bound states are
obtained by solving either the Schr\"odinger equation with a
Woods-Saxon potential~\cite{Cattapan00} or the Dirac equation with
self-consistent RMF potentials~\cite{Zhang04}.

The real stabilization method (RSM) is another bound-state-like
method~\cite{Hazi70}. The equation of motion of the system in
question is solved in a basis~\cite{Hazi70} or a box~\cite{Maier80}
of finite sizes, thus a bound state problem being always imposed.
The RSM uses the fact that the energy of a ``resonant'' state is
stable against changes of the sizes of the basis or the box. It has
been used to calculate the resonance parameters in elastic and
inelastic scattering processes~\cite{Fels71, Fels72}. Some efforts
have also been made in order to calculate more efficiently resonance
parameters with the RSM~\cite{Taylor76, Mandelshtam93,
Mandelshtam94, Kruppa99}. In this work, we investigate single
particle resonances in atomic nuclei by combining the RSM and the
relativistic mean field (RMF) model.

The paper is organized as follows. In Sec.~\ref{sec:theory} we give
briefly the formalism for the RSM and the RMF model. The numerical
details, the results for $^{120}$Sn and discussions are given in
Sec.~\ref{sec:results}. Finally we give a brief summary.

\section{Formalism of the RMF model and the RSM}
\label{sec:theory}

\subsection{The relativistic mean field model}

The basic ansatz of the relativistic mean field (RMF) model is a
Lagrangian density where nucleons are described as Dirac spinors
which interact via the exchange of several mesons ($\sigma$,
$\omega$, and $\rho$) and the photon~\cite{Serot86, Reinhard89,
Ring96, Vretenar05, Meng06},
\begin{eqnarray}
\displaystyle
 {\cal L}
   & = &
     \bar\psi_i \left( i\rlap{/}\partial -M \right) \psi_i
    + \frac{1}{2} \partial_\mu \sigma \partial^\mu \sigma
    - U(\sigma)
    - g_{\sigma} \bar\psi_i \sigma \psi_i
   \nonumber \\
   &   & \mbox{}
    - \frac{1}{4} \Omega_{\mu\nu} \Omega^{\mu\nu}
    + \frac{1}{2} m_\omega^2 \omega_\mu \omega^\mu
    - g_{\omega} \bar\psi_i \rlap{/}{\mbox{\boldmath$\omega$}} \psi_i
   \nonumber \\
   &   & \mbox{}
    - \frac{1}{4} \vec{R}_{\mu\nu} \vec{R}^{\mu\nu}
    + \frac{1}{2} m_{\rho}^{2} \vec{\rho}_\mu \vec{\rho}^\mu
    - g_{\rho} \bar\psi_i \rlap{/} \vec{{\mbox{\boldmath$\rho$}}} \vec{\tau} \psi_i
   \nonumber \\
   &   &\mbox{}
    - \frac{1}{4} F_{\mu\nu} F^{\mu\nu}
    - e \bar\psi_i \frac{1-\tau_3}{2}\rlap{/}{\bf A} \psi_i ,
\label{eq:Lagrangian}
\end{eqnarray}
where the summation convention is used and the summation over $i$
runs over all nucleons, $\rlap{/} x \equiv \gamma^\mu x_\mu =
\gamma_\mu x^\mu$, $M$ the nucleon mass, and $m_\sigma$, $g_\sigma$,
$m_\omega$, $g_\omega$, $m_\rho$, $g_\rho$ masses and coupling
constants of the respective mesons. The nonlinear self-coupling for
the scalar mesons is given by~\cite{Boguta77}
\begin{equation}
   U(\sigma) = \dfrac{1}{2} m^2_\sigma \sigma^2
              +\dfrac{g_2}{3}\sigma^3 + \dfrac{g_3}{4}\sigma^4 ,
\end{equation}
and field tensors for the vector mesons and the photon fields are
defined as
\begin{eqnarray}
 \left\{
  \begin{array}{rcl}
   \Omega_{\mu\nu}  & = & \partial_\mu\omega_\nu
                         -\partial_\nu\omega_\mu, \\
   \vec{R}_{\mu\nu} & = & \partial_\mu\vec{\rho}_\nu
                         -\partial_\nu\vec{\rho}_\mu
                         -g_{\rho} (\vec{\rho}_\mu
                                    \tm \vec{\rho}_\nu ), \\
   F_{\mu\nu}       & = & \partial_\mu {A}_\nu
                         - \partial_\nu {A}_\mu.
  \end{array}
 \right.
 \label{eq:tensors}
\end{eqnarray}

The classical variation principle gives equations of motion for the
nucleon, mesons and the photon. As in many applications, we study
the ground state properties of nuclei with time reversal symmetry,
thus the nucleon spinors are the eigenvectors of the stationary
Dirac equation
\begin{equation}
  \left[ \bm{\alpha} \cdot \bm{p} + V(\bm{r}) + \beta (M + S(\bm{r}))
  \right] \psi_i(\bm{r}) = \epsilon_i \psi_i(\bm{r}) ,
\label{eq:Dirac0}
\end{equation}
and equations of motion for mesons and the photon are
\begin{eqnarray}
 \left\{
   \begin{array}{rcl}
    \left( -\Delta + \partial_\sigma U(\sigma) \right )\sigma(\bm{r})
      & = & -g_\sigma \rho_s(\bm{r}) , \\
    \left( -\Delta + m_\omega^2 \right )             \omega^0(\bm{r})
      & = &  g_\omega \rho_v(\bm{r}) , \\
    \left( -\Delta + m_\rho^2 \right)                  \rho^0(\bm{r})
      & = &  g_\rho   \rho_3(\bm{r}) , \\
    -\Delta                                               A^0(\bm{r})
      & = &  e        \rho_p(\bm{r}) ,
   \end{array}
 \right.
 \label{eq:mesonmotion}
\end{eqnarray}
where $\omega^0$ and $A^0$ are time-like components of the vector
$\omega$ and the photon fields and $\rho^0$ the 3-component of the
time-like component of the iso-vector vector $\rho$ meson.
Equations~(\ref{eq:Dirac0}) and (\ref{eq:mesonmotion}) are coupled
to each other by the vector and scalar potentials
\begin{eqnarray}
 \left\{
   \begin{array}{lll}
     V(\bm{r}) & = & g_\omega \omega^0(\bm{r})
                    +g_\rho \tau_3 \rho^0(\bm{r})
                    +e \dfrac{1-\tau_3}{2} A^0(\bm{r}) , \\
     S(\bm{r}) & = & g_\sigma \sigma(\bm{r}), \\
   \end{array}
 \right.
 \label{eq:vaspot}
\end{eqnarray}
and various densities
\begin{eqnarray}
 \left\{
  \begin{array}{rcl}
   \rho_s(\bm{r})
   & = &
    \sum_{i=1}^A \bar\psi_i(\bm{r}) \psi_i(\bm{r}) ,\\
   \rho_v(\bm{r})
   & = &
    \sum_{i=1}^A \psi_i^\dagger(\bm{r}) \psi_i(\bm{r}) ,\\
   \rho_3(\bm{r})
   & = &
    \sum_{i=1}^A \psi_i^\dagger(\bm{r}) \tau_3 \psi_i(\bm{r}) ,\\
   \rho_c(\bm{r})
   & = &
    \sum_{i=1}^A \psi_i^\dagger(\bm{r})
                 \dfrac{1-\tau_3}{2}\psi_i(\bm{r}) .
  \end{array}
 \right.
 \label{eq:mesonsource}
\end{eqnarray}

For spherical nuclei, meson fields and densities depend only on the
radial coordinate $r$, the Dirac spinor reads
\begin{equation}
 \psi_{\alpha\kappa m}(\bm{r},s,t) =
   \left(
     \begin{array}{c}
       i \dfrac{G_\alpha^{\kappa}(r)}{r} Y^l _{jm} (\theta,\phi,s)
       \\
       - \dfrac{F_\alpha^{\kappa}(r)}{r} Y^{\tilde l}_{jm}(\theta,\phi,s)
     \end{array}
   \right) \chi_{t_\alpha}(t),
   \ \ j = l\pm\frac{1}{2},
 \label{eq:SRHspinor}
\end{equation}
with $Y^l _{jm}(\theta,\phi)$ the spin spherical harmonics. The
radial equation of the Dirac spinor, Eq. (\ref{eq:Dirac0}), is
reduced as
\begin{equation}
 \left\{
   \begin{array}{lll}
    \epsilon_\alpha G_\alpha^{\kappa} & = &
     \left( -\dfrac{\partial}{\partial r} + \dfrac{\kappa}{r}
     \right) F_\alpha^{\kappa}
     + \left( M + S(r) + V(r) \right) G_\alpha^{\kappa} ,
    \\
    \epsilon_\alpha F_\alpha^{\kappa} & = &
     \left( +\dfrac{\partial}{\partial r} + \dfrac{\kappa}{r}
     \right) G_\alpha^{\kappa}
     - \left( M + S(r) - V(r) \right) F_\alpha^{\kappa} .
   \end{array}
 \right.
 \label{eq:SRHDirac}
\end{equation}
The meson field equations become simply radial Laplace equations of
the form
\begin{equation}
 \left( - \frac{\partial^2}{\partial r^2}
        - \frac{2}{r}\frac{\partial}{\partial r} + m_{\phi}^2
 \right) \phi(r)
 = s_{\phi}(r).
 \label{eq:SRHmesonmotion}
\end{equation}
$m_{\phi}$ are the meson masses for $\phi = \sigma, \omega,\rho$ and
zero for the photon. The source terms are
\begin{eqnarray}
 s_{\phi}(r) =
  \left\{
    \begin{array}{ll}
      - g_\sigma\rho_s(r) - g_2 \sigma^2(r)  - g_3 \sigma^3(r),
     & \text{for}\ \sigma, \\
        g_\omega \rho_v(r),
     & \text{for}\ \omega, \\
        g_{\rho} \rho_3(r),
     & \text{for}\ \rho, \\
        e \rho_c(r),
     & \text{for}\ A, \\
    \end{array}
  \right.
 \label{eq:sources}
\end{eqnarray}
with
\begin{eqnarray}
 \left\{
  \begin{array}{lll}
   4\pi r^2 \rho_s(r) & = & \sum_{i=1}^A (|G_i(r)|^2 - |F_i(r)|^2), \\
   4\pi r^2 \rho_v(r) & = & \sum_{i=1}^A (|G_i(r)|^2 + |F_i(r)|^2), \\
   4\pi r^2 \rho_3(r) & = & \sum_{i=1}^A 2t_i
                                         (|G_i(r)|^2 + |F_i(r)|^2), \\
   4\pi r^2 \rho_c(r) & = & \sum_{i=1}^A \left(\frac{1}{2}-t_i\right)
                                         (|G_i(r)|^2 + |F_i(r)|^2). \\
  \end{array}
 \right.
 \label{eq:mesonsourceS}
\end{eqnarray}
The above coupled equations can be solved iteratively in $r$
space~\cite{Horowitz81} or in the harmonic oscillator
basis~\cite{Gambhir90} using the no sea and the mean field
approximations.

\subsection{The real stabilization method in coordinate space}

With the self consistent vector and scalar potentials $V(r)$ and
$S(r)$, the Dirac equation (\ref{eq:SRHDirac}) is solved in a
spherical box of the size $R_\mathrm{max}$ under the box boundary
condition, and thus the continuum is discretized. When
$R_\mathrm{max}$ is large enough, the energy of a bound state does
not change with $R_\mathrm{max}$. In the continuum region, there are
some states stable against the size of the box, i.e., the energy of
each of such states is almost constant with changing
$R_\mathrm{max}$; such stable states correspond to resonances.

The resonant parameters, $E_\gamma$ and $\Gamma$, may be obtained by
fitting the energy $E$ and the phase shift $\eta(E)$ in an energy
range around a resonance to the following formula~\cite{Hazi70},
\begin{equation}
 \label{eq:E-eta}
 \eta_l(E) = \eta_{l,\mathrm{pot}}(E)
         + \tan^{-1}\left( \frac{\Gamma/2}{E-E_\gamma} \right)
 .
\end{equation}
The phase shift $\eta_l(E)$ can be calculated as~\cite{Zhang07}
\begin{equation}
 \tan\left( \eta_l-\frac{l\pi}{2} \right)
 = - \frac{ \int^{R_\mathrm{max}}_0 \chi_l(r) \left[E-H(r)\right]f(r) \sin kr dr }
          { \int^{R_\mathrm{max}}_0 \chi_l(r) \left[E-H(r)\right]f(r) \cos kr dr }
 ,
 \label{eq:eta1}
\end{equation}
with $f(r)$ satisfying $f(r) \rightarrow 1$ when $r \rightarrow
\infty$ and $f(0)=f'(0)=0$. However, Eq.~(\ref{eq:eta1}) converges
very slowly with the box size due to the influence of the non zero
centrifugal potential at large ${R_\mathrm{max}}$~\cite{Zhang07}.

In the present work, we use a simpler method proposed by Maier et
al.~\cite{Maier80} in which it's not necessary to calculate the
phase shift. The resonance energy is determined by the condition
$\partial^2 E /
\partial R^2_\mathrm{max}$ = 0 and the corresponding box size is
labeled as $\bar R_\mathrm{max}$, i.e., $E_\gamma=E(\bar
R_\mathrm{max})$. The width is evaluated from the stability behavior
of the positive energy state against the box size around $\bar
R_\mathrm{max}$.

When $r$ is large enough, the nuclear potentials $S(r)$ and $V(r)$
vanish, $G(r)/r$ satisfy
\begin{equation}
   \displaystyle
    \frac{d^2G}{dr^2}+\left(\alpha^2-\frac{\kappa(\kappa+1)}{r^2}\right)G=0,
\end{equation}
with $\alpha^2=E^2-M^2$. The general solution reads
\begin{equation}
  G(r) \propto \alpha r\left[\cos\eta_l\ j_l(\alpha r)- \sin\eta_l\ n_l(\alpha r)\right], 
 \label{eq:scattering}
\end{equation}
When $r\rightarrow \infty$, $G(r) \propto \sin(\alpha r -
\frac{l\pi}{2} + \eta_l)$. Therefore when $R_\mathrm{max}$ is large
enough,
\begin{equation}
 \alpha R_\mathrm{max} - \frac{l\pi}{2} + \eta_l = n\pi .
 \label{eq:node}
\end{equation}
Under the assumption that the phase shift from the potential
scattering $\eta_{l,\mathrm{pot}}(E)$ varies slowly with respect to
the box size, i.e, $\partial\eta_{l,\mathrm{pot}}/\partial
R_\mathrm{max} \sim 0$, one derives from Eqs.~(\ref{eq:E-eta}) and
(\ref{eq:node}) the formula,
\begin{equation}
 \Gamma =
 \frac{2\sqrt{E_\gamma^2+2E_\gamma M}}
      {-(E_\gamma+M)\bar R_\mathrm{max}-(E_\gamma^2+2E_\gamma M)\left[\partial E/\partial R_\mathrm{max}|_{\bar R_\mathrm{max}}\right]^{-1}}
 \label{eq:Gamma}
 .
\end{equation}
In the non-relativistic limit, $E_\gamma\ll M$, Eq.~(\ref{eq:Gamma})
is reduced to
\begin{equation}
 \Gamma =
 \frac{2\sqrt{2E_\gamma/M}}
      {-\bar R_\mathrm{max}-2E_\gamma\left[dE/dR_\mathrm{max}|_{\bar R_\mathrm{max}}\right]^{-1}}
 \label{eq:Gamma2}
 ,
\end{equation}
which is essentially the same as Eq.~(8) in Ref.~\cite{Maier80}
except for that here natural units with $\hbar=c=1$ is used.

\section{Results and Disscusion}
\label{sec:results}

\begin{figure}
\includegraphics[width=8cm]{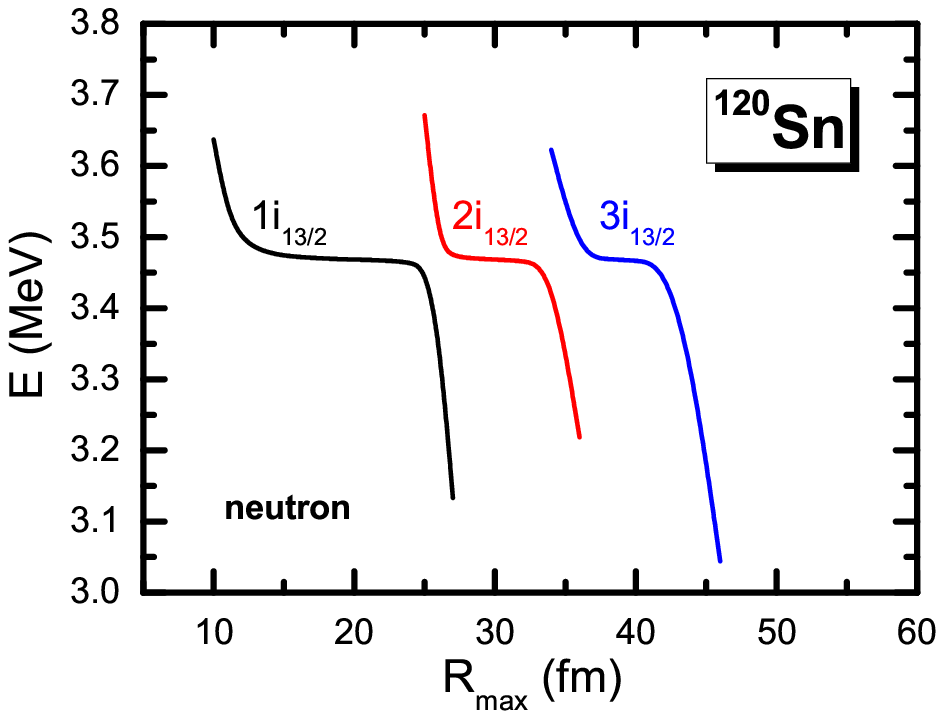}
\caption{\label{fig:i13/2}Positive energy $\nu$i$_{13/2}$ states in
$^{120}$Sn under different box boundary conditions.}
\end{figure}

\begin{figure}
\includegraphics[width=8cm]{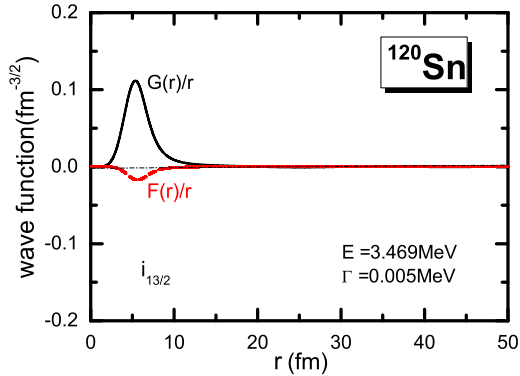}
\caption{\label{fig:i13/2-wf}The radial wave function for the
neutron $\nu$i$_{13/2}$ resonant state in $^{120}$Sn.}
\end{figure}

In this section we present the results of the RSM in the framework
of the RMF model. In our calculation, we use for the Lagrangian
density the effective interactions PK1~\cite{Long04} and
NL3~\cite{Lalazissis97}. We change the size of the box in a large
range (7 fm $< R_\mathrm{max} <$ 60 fm if not specified) in order to
find not only narrow resonances but also wide ones. We take
$^{120}$Sn as an example and compare the results for neutron
resonances from the RSM with those from the ACCC
approach~\cite{Zhang04} and the scattering phase shift
method~\cite{Sandulescu03}.

By examining the stability of low lying positive energy states
against the size of the box $R_\mathrm{max}$, we find that except
for s state, there are neutron resonances in $^{120}$Sn with the
orbital angular momentum $l$ up to 6.

The narrowest resonance is an i$_{13/2}$ state which lies at about
3.45 MeV above the threshold. The positive energy $\nu$i$_{13/2}$
states in a box of different sizes are shown in
Fig.~\ref{fig:i13/2}. With the box size $R_\mathrm{max}$ increasing,
the lowest $\nu$i$_{13/2}$ state first falls down quickly then its
energy becomes constant in a large region of $R_\mathrm{max}$. After
it crosses with the second lowest $\nu$i$_{13/2}$ state at around
$R_\mathrm{max} = 25$ fm, the lowest one falls down again. Similar
level-crossings occur regularly at larger $R_\mathrm{max}$. This
stability behavior implies that there is a narrow $\nu$i$_{13/2}$
resonant state with approximate energy 3.45 MeV. The resonant energy
3.469 MeV and $\bar R_\mathrm{max} = 20.1$ fm are obtained under the
condition $\partial^2 E /
\partial R^2_\mathrm{max}$ = 0 from the first $E\sim R_\mathrm{max}$
curve (labeled as ``1i$_{13/2}$'') in Fig.~\ref{fig:i13/2}. The
width 0.003 MeV are obtained from Eq.~(\ref{eq:Gamma}). One of the
approximations made in deriving Eq.~(\ref{eq:Gamma}) is
$R_\mathrm{max}$ should be large. We next examine the dependence of
the resonant parameters on the box size by calculating $E_\gamma$
and $\Gamma$ from other $E\sim R_\mathrm{max}$ curves with larger
$R_\mathrm{max}$. For this purpose the calculations with
$R_\mathrm{max}$ up to 65 fm are carried out. The results are given
in Table~\ref{tab:conv}. The energy is almost a constant with
increasing $\bar R_\mathrm{max}$. The variation between the widths
obtained from adjacent $E\sim R_\mathrm{max}$ curves decreases with
$\bar R_\mathrm{max}$ and is about 1\% at $\bar R_\mathrm{max}\sim
63$ fm. For other resonant states presented in this work, we also
make similar investigations. Once it converges to within 1\%, the
value of the width is assigned to a resonant state. The wave
function of the resonant state $\nu$i$_{13/2}$ is given in
Fig.~\ref{fig:i13/2-wf}. From this figure one can find that this
state is almost localized inside the nucleus which is consistent
with the small width $\Gamma$.

\begin{table}
\caption{\label{tab:conv}Energies and widths of the single neutron
resonant state i$_{13/2}$ in $^{120}$Sn from different $E\sim
R_\mathrm{max}$ curves (cf. Fig.~\protect\ref{fig:i13/2}). $\bar
R_\mathrm{max}$ is in fm and $E_\gamma$ and $\Gamma$ are in MeV.}
\begin{tabular}{c|c|c||c|c|c}
\hline\hline
 $\bar R_\mathrm{max}$ & $E_\gamma$ & 100$\times\Gamma$ &
 $\bar R_\mathrm{max}$ & $E_\gamma$ & 100$\times\Gamma$ \\
\hline
 20.1 & 3.4688 & 0.327 &
 47.2 & 3.4686 & 0.470 \\
 30.3 & 3.4686 & 0.426 &
 55.3 & 3.4685 & 0.478 \\
 38.9 & 3.4686 & 0.456 &
 63.2 & 3.4686 & 0.483 \\
\hline\hline
\end{tabular}
\end{table}

\begin{figure}
\includegraphics[width=8cm]{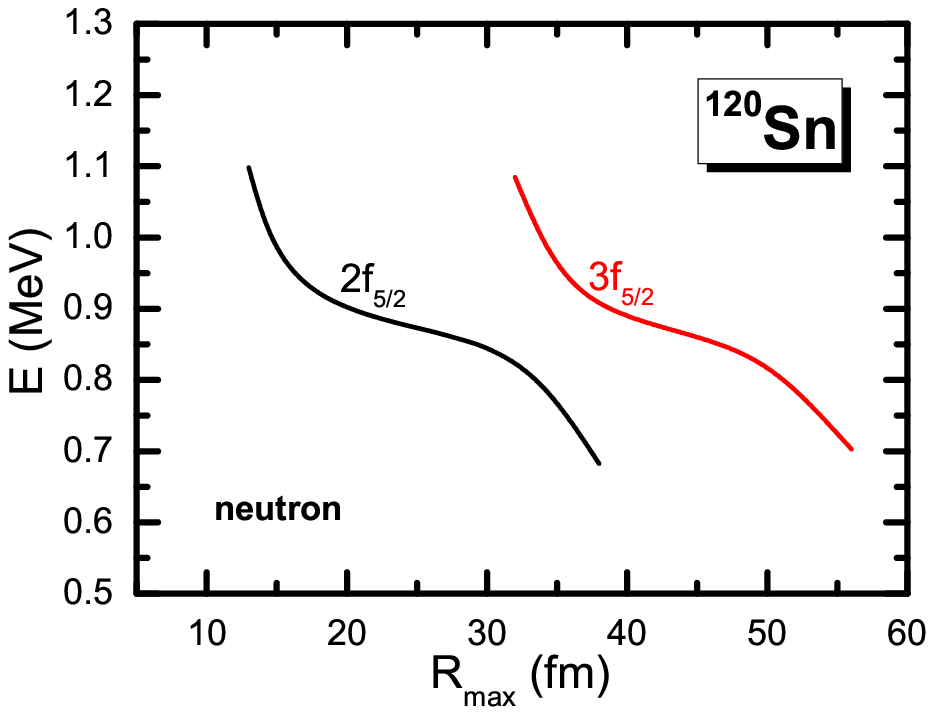}
\caption{\label{fig:f5/2}Positive energy $\nu$f$_{5/2}$ states in
$^{120}$Sn under different box boundary conditions.}
\end{figure}

\begin{figure}
\includegraphics[width=8cm]{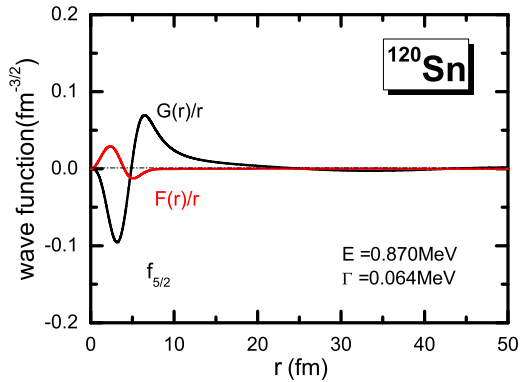}\\
\caption{\label{fig:f5/2-wf}The radial wave function for the neutron
$\nu$f$_{5/2}$ resonant state in $^{120}$Sn.}
\end{figure}

Although it lies below the state $\nu$i$_{13/2}$, the resonant state
$\nu$f$_{5/2}$ is about an order of magnitude wider than
$\nu$i$_{13/2}$. The reason is that the centrifugal barrier for
$\nu$f$_{5/2}$ ($l=3$) is much lower than that for $\nu$i$_{13/2}$
($l=6$). The stability behavior of the state $\nu$f$_{5/2}$ is
presented in Fig.~\ref{fig:f5/2}. The resonant energy and width are
0.870 MeV and 0.064 MeV, respectively. The wave function of this
state is also well localized as shown in Fig.~\ref{fig:f5/2-wf}.

\begin{figure}
\includegraphics[width=8cm]{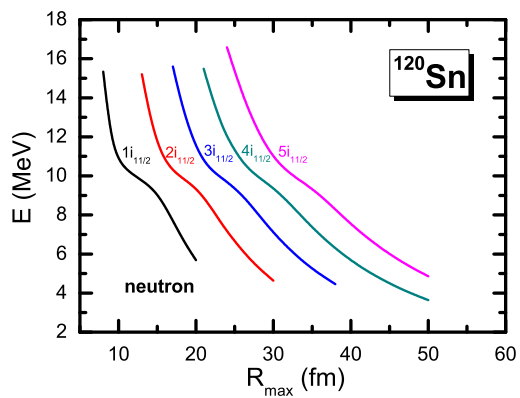}\\
\caption{\label{fig:i11/2}Positive energy $\nu$i$_{11/2}$ states in
$^{120}$Sn under different box boundary conditions.}
\end{figure}

\begin{figure}
\includegraphics[width=8cm]{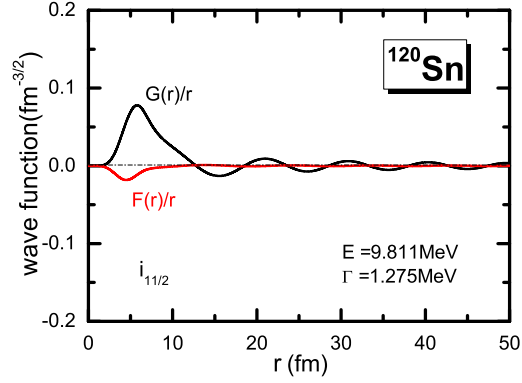}
\caption{\label{fig:i11/2-wf}The radial wave function for the
neutron $\nu$i$_{11/2}$ resonant state in $^{120}$Sn.}
\end{figure}

Except the state $\nu$i$_{13/2}$, there is another neutron resonant
state with $l=6$, i.e., the $\nu$i$_{11/2}$ state. As shown in
Fig.~\ref{fig:i11/2}, there is a less stable state lying at about 10
MeV. Although it shares the same centrifugal barrier with
$\nu$i$_{13/2}$, $\nu$i$_{11/2}$ is much wider because its energy is
much larger than that of $\nu$i$_{13/2}$. In
Fig.~\ref{fig:i11/2-wf}, one finds also that the wave function of
$\nu$i$_{13/2}$ oscillates very much even at $r=50$ fm. The
resonance parameters for this state are $E_\gamma=9.811$ MeV and
$\Gamma=1.275$ MeV respectively.

\begin{figure}
\includegraphics[width=8cm]{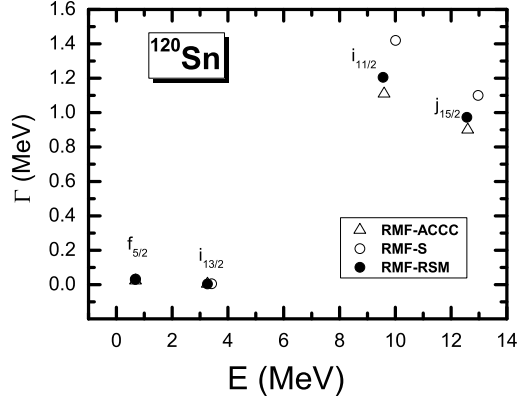}
\caption{\label{fig:ewcompare}Energies and widths of single neutron
resonant states in $^{120}$Sn from different methods. RMF-ACCC,
RMF-RSM, and RMF-S represents results from the analytical
continuation in the coupling constant approach, the real
stabilization method, and the scattering phase shift method in the
framework of the RMF model with NL3 parameter set. }
\end{figure}

\begin{table}
\caption{\label{tab:ewcompare}Energies and widths of single neutron
resonant states in $^{120}$Sn from different methods. RMF-ACCC,
RMF-RSM, and RMF-S represents results from the analytical
continuation in the coupling constant approach, the real
stabilization method, and the scattering phase shift method in the
framework of the RMF model. In the RMF-RSM calculations, both PK1
and NL3 parameter sets are used. All quantities are in MeV.}
\begin{tabular}{l|cc|cc|cc|cc}
\hline\hline
         & \multicolumn{2}{c}{RMF-RSM (PK1)}  & \multicolumn{2}{c}{RMF-RSM (NL3)}
         & \multicolumn{2}{c}{RMF-ACCC (NL3)} & \multicolumn{2}{c}{RMF-S (NL3)}   \\
 $\nu l_j$       & $E_\gamma$& $\Gamma$ & $E_\gamma$& $\Gamma$  & $E_\gamma$& $\Gamma$ & $E_\gamma$  & $\Gamma$  \\
\hline
 $\nu$f$_{5/2}$  & 0.870 & 0.064 &  0.674 & 0.030 &  0.685 & 0.023 &  0.688 & 0.032 \\
 $\nu$i$_{13/2}$ & 3.469 & 0.005 &  3.266 & 0.004 &  3.262 & 0.004 &  3.416 & 0.005 \\
 $\nu$i$_{11/2}$ & 9.811 & 1.275 &  9.559 & 1.205 &  9.60  & 1.11  & 10.01  & 1.42  \\
 $\nu$j$_{15/2}$ &12.865 & 1.027 & 12.564 & 0.973 & 12.60  & 0.90  & 12.97  & 1.10  \\
\hline\hline
\end{tabular}
\end{table}

The energies and widths of single particle neutron resonant states
obtained from the RSM calculations are summarized in
Table~\ref{tab:ewcompare}. We also calculate these resonances using
the parameter set NL3 for the Lagrangian density in the RMF model
and compare the present results with those from the ACCC approach
and the scattering phase shift method~\cite{Zhang04b} in
Table~\ref{tab:ewcompare}. In Fig.~\ref{fig:ewcompare}, the
comparison is also made in a planar $E_\gamma$-$\Gamma$ plot. For
the two low lying resonant states, $\nu$f$_{5/2}$ and
$\nu$i$_{13/2}$, the three methods give consistent results both for
the energy and the width. Although similar energies are obtained
from these three models for higher resonances, $\nu$i$_{11/2}$ and
$\nu$j$_{15/2}$, clear differences occur among the widths from the
ACCC approach and the scattering method and the results from the RSM
lies in between.

\section{Summary}
\label{sec:summary}

In summary, the real stabilization method (RSM) has been developed
within the framework of the relativistic mean field (RMF) model.
With the self-consistent nuclear potentials provided by the RMF
calculations with the parameter sets PK1 and NL3 for the Lagrangian
density, the Dirac equation for the neutron is solved in the
coordinate space under the box boundary condition. By investigating
the stable behavior of the positive energy states against changes of
the box size, the resonant states are singled out. The RMF-RSM is
used to study single neutron resonant states in spherical nuclei. As
examples, the energies, widths and wave functions of low-lying
neutron resonant states in $^{120}$Sn are obtained. Since a very
large box size is used, even wider resonances can also be found.
These results are compared with those from the scattering phase
shift method and the analytic continuation in the coupling constant
method and satisfactory agreements are found.

\begin{acknowledgments}
Helpful discussions with Lisheng Geng, Zhipan Li, and Hongfeng
L\"{u} are acknowledged. This work was partly supported by the
National Natural Science Foundation of China under Grant Nos.
10435010, 10475003, and 10575036, the Major State Basic Research
Development Program of China under contract No. 2007CB815000 and the
Knowledge Innovation Project of Chinese Academy of Sciences under
contract Nos. KJCX-SYW-N2 and KJCX2-SW-N17. Part of the computation
of this work was performed on the HP-SC45 Sigma-X parallel computer
of ITP and ICTS and supported by Supercomputing Center, CNIC, CAS.
\end{acknowledgments}

\end{document}